# Dissipative Solitons: Perturbations and Chaos Formation

## Vladimir L. Kalashnikov

Instutute for Photonics, Technical University of Vienna, Vienna, Austria
Email: kalashnikov@tuwien.ac.at

**Abstract:** The perturbations of chirped dissipative solitons are analyzed in the spectral domain. It is shown, that the structure of the perturbed chirped dissipative soliton is highly nontrivial and has a tendency to an enhancement of the spectral perturbations especially at the spectrum edges, where the irregularities develop. Even spectrally localized perturbations spread over a whole soliton spectrum. As a result of spectral irregularity, the chaotic dynamics develops due to the spectral loss action. In particular, the dissipative soliton can become fragmented though remains localized.

## 1. Introduction

The complex nonlinear Ginzburg-Landau equation (CGLE) has so broad horizon of application, that the concept of "the word of the Ginzburg-Landau equation" [1] has become broadly established. The CGLE demonstrates its effectiveness in quantum optics, condensate-matter physics, study of non-equilibrium phenomena and nonlinear dynamics, etc. In optics and laser physics, the CGLE provides an adequate description of ultra-short pulse formation and dynamics [2,3].

The CGLE is multiparameter and not integrable in a general form. Hence, a study of its solutions needs extensive numerical simulations. The exact analytical solutions are known for a few of cases, when they represent the solitary waves (dissipative solitons) [2]. As a rule, one presumes some class of functional expressions to construct such a restricted class of solutions. As a result, the solutions beyond a given class are missed. Moreover, the issues of stability and dynamical properties of the analyzed dissipative solitons remain open. These challenges stimulate interest in the approximate methods of integration of the CGLE [4] and the study of the dissipative soliton perturbations. In many cases, an excitation of such perturbations can result in the chaotic soliton dynamics [5,6].

A physically important sector, which permits an approximate analysis, is represented by the chirped dissipative solitons (CDSs) of the CGLE. The very important property of CDS is that such a soliton is energy scalable [7]. This allows obtaining the femtosecond laser pulses with about of and over microjoule energies [8]. A large chirp causes the spectral extra-broadening of CDS and an appearance of spectral perturbations [5,6,9]. As a result, the chaotic dynamics of CDS (so-called, chaotic mode locking) can develop.

In this work, I present the extension of the approximate analytical technique [4,10,11,12] to analysis of CDS of the perturbed complex nonlinear CGLE. Three main sources of perturbations will be considered: i) higher-order (quintic) nonlinearity, ii) higher-order dispersions, and iii) narrowband absorption. As will be shown, the perturbations under consideration distort the CDS spectrum that affects the soliton dynamics and, in particular, leads to the chaos appearance.

## 2. Perturbative analysis in spectral domain

To be obeying the CGLE, the electromagnetic field with the amplitude $A(z,t)$ ($t$ is the local time, and $z$ is the propagation coordinate) has to be described on basis of the slowly varying amplitude approximation, that is provided by the relation $\omega_0 \gg 1/T$ ($\omega_0$ is the field carrier frequency, and $T$ is the soliton width). When one can neglect the field variation along a cavity round-trip (for an oscillator) or the variation of material parameters along a fiber (for intra-fiber propagation), the field dynamics can be described on basis of the perturbed cubic CGLE [2]

$$\frac{\partial A(z,t)}{\partial z} = [-\sigma + (\kappa - i\gamma)P(z,t)]A(z,t) + \\ + (\alpha + i\beta)\frac{\partial^2}{\partial t^2}A(z,t) + \Gamma\{A(z,t)\}. \quad (1)$$

Here $P=|A|^2$ is the instant field power, $\alpha$ is the squared inverse gain bandwidth, $\beta$ is the group delay dispersion (GDD) parameter (the case of $\beta>0$ corresponds to a normal dispersion and is under consideration in this work), and $\sigma$ is the net-loss parameter. The nonlinear terms in Eq. (1) describe i) self-amplitude modulation with the parameter $\kappa$, and ii) self-phase modulation with the parameter $\gamma$. $\Gamma$ is the perturbation, which functional form depends on the type of perturbation (see below).

The unperturbed Eq. (1) has the exact CDS solution in the form [2]

$$A(z,t) = \sqrt{P_0}\,\text{sech}(t/T)^{1-i\psi}\exp(-iqz), \quad (2)$$

where $T$ is the soliton width, $P_0$ is the soliton peak power, $\psi$ is the chirp, and $q$ is the soliton wavenumber. However, this exact solution is not promissory for a further perturbative analysis because it results from the assumed restriction on the CDS phase profile and, as a result, the constraint on $\beta$. Moreover, such a form of solution has the Fourier image, which is expressed in terms of the beta function, that complicates an anlysis in the spectral domain. Therefore it is convenient to use the technique developed in [4,10-13]. This technique allows obtaining the approximate expression for the Fourier image of (2)

$$a(\omega) \approx \sqrt{6\pi\beta/\kappa(1+c)}\exp\left[3i\gamma\omega^2/2\kappa(1+c)(\Delta^2-\omega^2)\right]H(\Delta^2-\omega^2), \quad (3)$$

where $\omega$ is the deviation from $\omega_0$, $c=\alpha\gamma/\beta\kappa$, $\Delta^2=3\sigma c/\alpha(2-c)$, and H is the Heaviside function. Other CDS parameters are $\psi = 3\gamma/\kappa(1+c)$, $T = 3\gamma/\kappa\Delta(1+c)$, and $P_0 = \beta\Delta^2/\gamma$. The solution (3) can be easily generalized for the more complicated nonlinear terms in the CGLE [4,10,13]. Another advantage of the solution (3) is that it allows developing an analytical perturbative theory of the solitonic sector of Eq. (1) in the spectral domain.

The equation for the Fourier image $f(\omega)$ of the small perturbation of $a(\omega)$ can be written as [2]

$$[q-k(\omega)]f(\omega)+\frac{1}{\pi}\int_{-\infty}^{\infty}U(\omega-\omega')f(\omega')d\omega' + \\ +\frac{1}{2\pi}\int_{-\infty}^{\infty}V(\omega-\omega')f^*(\omega')d\omega' = S(\omega), \quad (4)$$

where the linear wave wavenumber is $k(\omega) = \beta\omega^2 + i(\sigma + \alpha\omega^2)$, the kernels $U$ and $V$ are the Fourier images of $|a(t)|^2$ and $a(t)^2$, respectively ($a(t)$ is the time-dependent part of (2)). The source term $S(\omega)$ is the Fourier image of $\Gamma\{a(t)\}$ multiplied by $i$. Further, one may assume a phase matching between the soliton and its perturbation. This assumption allows $U=V$ and Eq. (4) can be solved through the Neumann series

$$f_n(\omega) = \frac{S(\omega)}{q-k(\omega)} - \frac{3}{\pi[q-k(\omega)]}\int_{-\infty}^{\infty}U(\omega-\omega')f_{n-1}(\omega')d\omega', \quad (5)$$

where $f_n$ is the $n$-th iteration and $f_0 = S(\omega)/[q-k(\omega)]$. The applications of the considered technique will be considered in the next section.

## 3. Perturbations of CDS

Let us consider some basic perturbation affecting the CDS. It will be seen, that the spectral perturbations can become irregular that induces a chaotic dynamics due to a feedback caused by the spectral dissipation.

### 3.1. Quintic nonlinearity

Let the quintic nonlinear term in Eq. (1) be $\Gamma\{A(z,t)\} = -\kappa\zeta P(z,t)^2 A(z,t)$. The tabletop spectrum of exact numerical solution is shown in Fig. 1 by solid black curve. The analytical approximations are shown by dashed and gray curves. They are very close to the numerical one, but have abrupt edges due to the Heaviside function in (3). One can conclude that the quintic nonlinear term can be treated as a perturbation in the case under consideration.

The successive terms of (5) are shown in Fig. 2. One can see that the tabletop spectrum becomes convex and some oscillating substructure develops. The last phenomenon explains the results of [5,6], where the cause of both regular and chaotic pulsations of CDS has been attributed to an excitation of the solitonic internal modes.

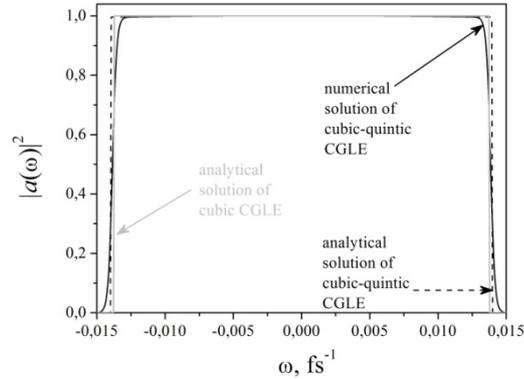

Figure 1. CDS spectra for i) numerical (solid black curve) and ii) approximate analytical solutions of the cubic-quintic CGLE (dashed black curve), as well as iii) the cubic CGLE (solid gray curve). Parameters correspond to a Cr:ZnSe mode-locked oscillator: $\alpha$=16 fs$^2$, $\beta$=250 fs$^2$, $\kappa$=0.04$\gamma$; $\zeta$=0.2$\gamma$ (black curves) and 0 (gray curve).

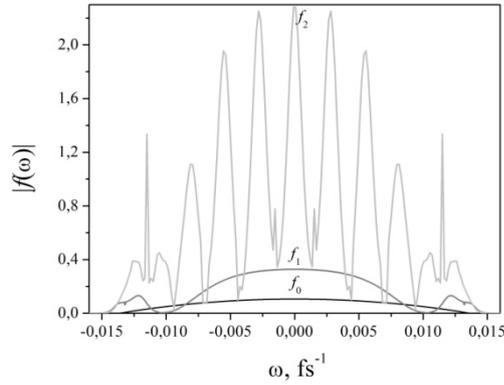

Figure 2. Absolute values of the successive terms in the Neumann series (5) for the quintic nonlinear perturbation.

### 3.2. Higher-order dispersions

Let us consider the third-order dispersion (TOD) term in Eq. (1), which can be expressed as $\Gamma\{A(z,t)\} = \theta \partial^3 A(z,t)/\partial t^3$ ($\theta$ is the TOD coefficient). The perturbed spectra (in zero- and first-orders of (5)) are presented in Fig. 3. An important feature of the perturbed CDS in presence of higher-order dispersion is an appearance of oscillating substructure growing to the spectrum edges. The phase changes dramatically in the vicinity of $\pm\Delta$ (Fig. 4) and this domain is very sensitive to perturbations due to their parametric amplification expressed by the term $1/[q - k(\omega)] \propto 1/(\Delta^2 - \omega^2)$ in Eq. (5). The induced substructure is fine and has a tendency to the growth of irregularities. On the other hand, the edges of spectrum are subjected to the maximum spectral loss described by the term $\alpha \partial^2/\partial t^2$ in Eq. (1). As a result of such loss, the spectral irregularities nearby the spectrum edges cause the irregular pulsations of soliton and, eventually, the chaotic mode-locking (Fig. 5), which has been reported in [6,9]. When the higher-order dispersions increase, the perturbations approach the spectrum center. Then the CDS becomes fragmented but remains localized (Fig. 5).

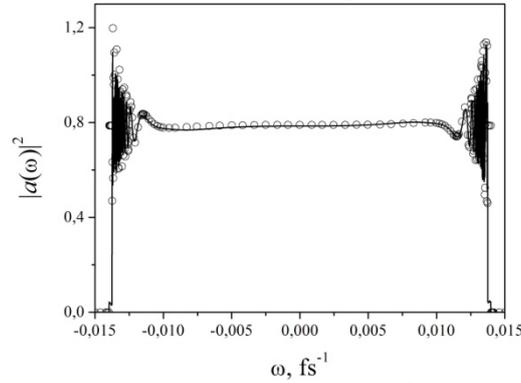

Figure 3. CDS spectra in presence of the TOD ($\theta$=300 fs$^3$) in zero- (curve) and first- (circles) orders of (5).

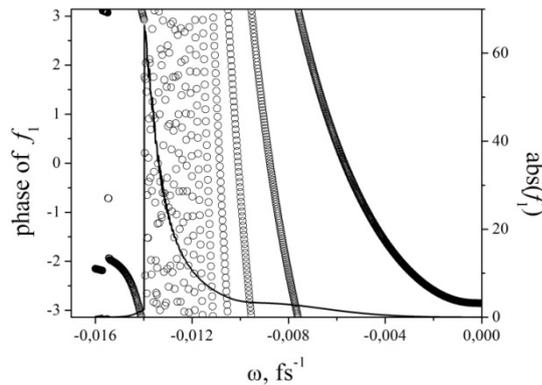

Figure 4. Absolute value (curve) and phase (circles) of $f_1$ corresponding to Fig. 3.

Another way to interpret the chaotic behavior in the presence of higher-order dispersions is to consider it as a result of the dispersive wave generation. Such a generation appears if the resonance condition is satisfied: $k(\omega) \equiv \beta\omega^2 + \sum_{l=3} \theta_l \omega^l = q.$ The resonant frequency $\omega_r$ providing such an condition is shifted inside the CDS spectrum (i.e. $|\omega_r|<\Delta$, Fig. 6) if the TOD parameter $\theta\neq 0$. The source term $S(\omega)$ for such a perturbation is $\propto \theta\omega^3 \text{H}(\Delta^2 - \omega^2)$ and increases towards the spectrum edges (Fig. 6) (unlike the case of the Schrödinger soliton [2]).

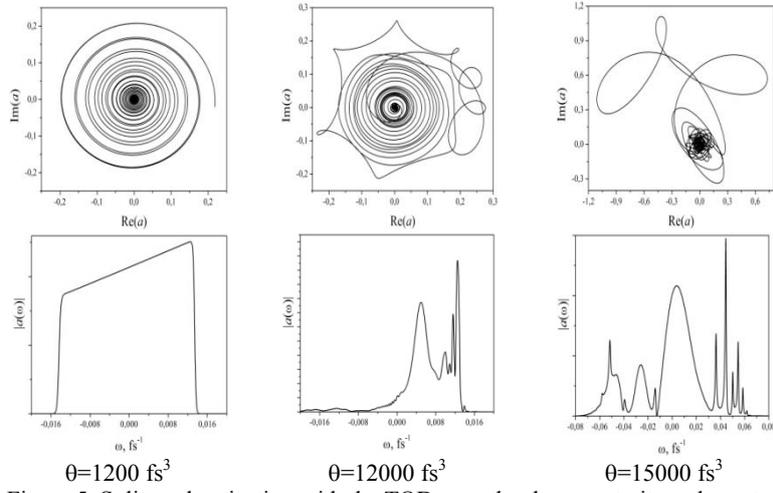

$\theta=1200$ fs$^3$  $\theta=12000$ fs$^3$  $\theta=15000$ fs$^3$

Figure 5. Soliton chaotization with the TOD growth: phase portraits and spectra.

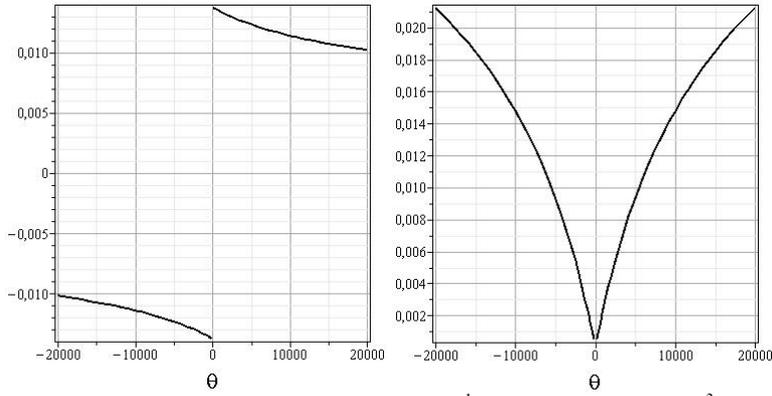

Figure 6. Left picture: resonant frequency $\omega_r$ [fs$^{-1}$] in dependence on $\theta$ [fs$^3$]; right picture: $\theta\omega_r^3$ in dependence on $\theta$.

### 3.3. Narrowband losses: a gas filled oscillator or an impured fiber

Let us consider the case of solitonic propagation within a medium filled with impurities consisting of the $l$ independent Lorentz lines centered at $\omega_l$ with the loss coefficients $\varepsilon_l<0$ and the linewidths $\Omega_l$. Then

$$S(\omega) = ia(\omega)\sum_l \varepsilon_l \frac{1-i(\omega-\omega_l)/\Omega_l}{1+[(\omega-\omega_l)/\Omega_l]^2} \quad . \tag{6}$$

For a single line centered at $\omega_1 = \omega_0$, the central part of the CDS spectrum is shown in Fig. 7 for $a(\omega)$ corresponding to the approximate solutions of the cubic and cubic-quintic CGLE perturbed by zero and first terms of the Neumann series (5). One can see, that the perturbed spectral profile reprodu-ces the real part of an absorber permittivity like the case of the Schrödinger soliton [14]. However, in contrast to the perturbed Schrödinger soliton, the CDS perturbation spreads along the soliton spectral profile (Fig. 8; parameters of Fig. 7). That is the perturbation is not strongly bounded in the vicinity of $\omega_l$ and the higher-order terms of the Neumann series contribute. As a result, the CDS becomes perturbed also through the spectral loss growing towards the spectrum edges as it was in the case of perturbations induced by higher-order dispersions (see above). Thus, the spectrally localized perturbation distorts a whole CDS spectrum.

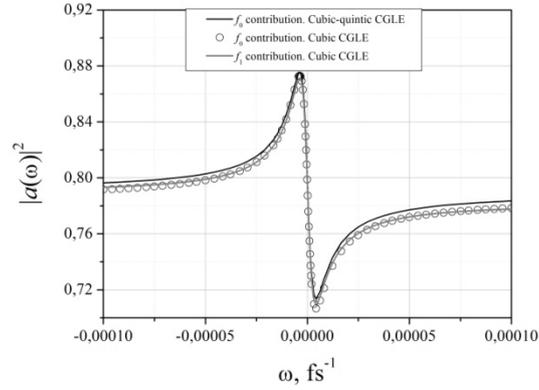

Figure 7. Central part of the CDS spectrum perturbed by a narrowband loss ($\Omega_1=4$ GHz, $\varepsilon_1=-0.005$).

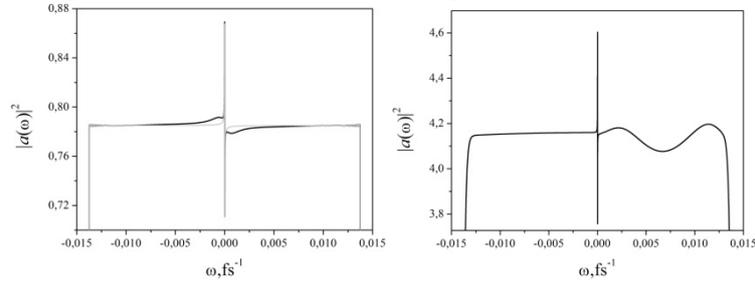

Figure 8. Left picture: the analytical CDS spectra (the cubic CGLE) perturbed by $f_0$ (gray) and $f_1$ (black). Right picture: the numerical spectrum (the cubic-quintic CGLE). $|a|^2$ is given in arbitrary units.

## 5. Conclusions

In this paper, the perturbative analysis of chirped dissipative solitons has been developed in the spectral domain. Three main types of perturbations have been considered: i) higher-order nonlinearity, ii) higher-order dispersions, and iii) narrowband losses. Unlike the perturbed Schrödinger soliton, the CDS has a nontrivial internal structure, which is strongly affected by perturbations. Even spectrally localized perturbations affect a whole CDS spectrum. Most sensitive parts of the CDS are its spectral edges, where the perturbations are resonantly enhanced and the irregularity develops. Since the spectral loss increases towards the spectrum edges, such a spectral irregularity becomes transformed into the irregular CDS dynamics.

**Acknowledgements** This work was supported by the Austrian Fonds zur Förderung der wissenschaftlichen Forschung, FWF project P20293.

## References

[1] I.S.Aranson, L.Kramer. The world of the complex Ginzburg-Landau equation. *Rev. Mod. Phys.*, 74:99, 2002.

[2] N.N.Akhmediev, A.Ankiewicz. *Solitons: Nonlinear Pulses and Beams*. Chapman&Hall, 1997.

[3] F.X.Kärtner, Ed. *Few-cycle Laser Pulse Generation and its Applications*. Springer Verlag, Berlin, 2004.

[4] V.L.Kalashnikov. Chirped dissipative solitons of the complex cubic-quintic nonlinear Ginzburg-Landau equation. *Phys. Rev. E*, 80:046606, 2009.

[5] V. L. Kalashnikov, A. Chernykh. Spectral anomalies and stability of chirped-pulse oscillators. *Phys. Rev. A*, 75:033820, 2007.

[6] V.L.Kalashnikov. Chaotic mode-locking of chirped-pulse oscillators, in *Proc. 2nd Chaotic Modeling and Simulation International Conference*, 2009.

[7] V.L.Kalashnikov, E.Podivilov, A.Chernykh, S.Naumov, A.Fernandez, R.Graf, A.Apolonski. Approaching the microjoule frontier with femtosecond laser oscillators. Theory and comparison with experiment. *New Journal of Physics*, 7:217, 2005.

[8] S.Naumov, A.Fernandez, R.Graf, P.Dombi, F.Krausz, A.Apolonski. Approaching the microjoule frontier with femtosecond laser oscillators. *New Journal of Physics*, 7:216, 2005.

[9] V.L.Kalashnikov, A.Fernández, A.Apolonski. High-order dispersion in chirped-pulse oscillators. *Optics Express*, 16:4206, 2008.

[10] E.Podivilov, V.L.Kalashnikov. Heavily-chirped solitary pulses in the normal dispersion region: new solutions of the cubic-quintic complex Ginzburg-Landau equation. *JETP Letters*, 82:467 (2005).

[11] V.L.Kalashnikov, E.Podivilov, A.Chernykh, A.Apolonski. Chirped-pulse oscillators: theory and experiment. *Applied Physics B*, 83:503, 2006.

[12] V.L.Kalashnikov, A. Apolonski. Chirped-pulse oscillators: A unified standpoint. *Phys. Rev. A*, 79:043829, 2009.

[13] V.L.Kalashnikov. The unified theory of chirped-pulse oscillators, in *Proc. SPIE*, M. Bertolotti, Ed., 7354:73540T, 2009.

[14] V.L.Kalashnikov, E.Sorokin, J.Mandon, G.Guelachvili, N.Picque, I.T.Sorokina. Femtosecond lasers for intracavity molecular spectroscopy, in *Conference Abstract Vol. 3rd EPS-QEOD Europhoton Conf. on Solid-State,Fiber and Waveguided Light Sources*, 32G: TUoA.3, 2008.